\documentclass[12pt,preprint]{aastex}
\usepackage{epsf}
\newcommand{\Msun}{\mathrm{M}_{\odot}}

\newcommand\loh{{12+\log{\rm(O/H)}}}

\begin{document}

\shortauthors{Prieto, Stanek, \& Beacom}

\title{Characterizing Supernova Progenitors via the Metallicities of
their Host Galaxies, from Poor Dwarfs to Rich Spirals}

\author{Jos\'e~L.~Prieto\altaffilmark{1}, Krzysztof~Z.~Stanek\altaffilmark{1,3},
John~F.~Beacom\altaffilmark{2,1,3}}

\altaffiltext{1}{\small Department of Astronomy, Ohio State University, Columbus, OH 43210 }
\altaffiltext{2}{\small Department of Physics, Ohio State University, Columbus, OH 43210}
\altaffiltext{3}{\small Center for Cosmology and Astro-Particle Physics, Ohio State University, Columbus, OH 43210}

\email{prieto@astronomy.ohio-state.edu}

\begin{abstract}
We investigate how the different types of supernovae are relatively
affected by the metallicity of their host galaxy. We match the SAI
supernova catalog to the SDSS-DR4 catalog of star-forming galaxies with
measured metallicities. These supernova host galaxies span a range of
oxygen abundance from $\loh=7.9$ to $9.3$ ($\sim 0.1$ to 2.7 solar) and
a range in absolute magnitude from M$_{B} = -15.2$ to $-22.2$. To reduce
the various observational biases, we select a subsample of
well-characterized supernovae in the redshift range from 0.01 to 0.04,
which leaves us with 58 SN~II, 19~Ib/c, and 38~Ia.  We find strong
evidence that SN~Ib/c are occurring in higher-metallicity host galaxies
than SN~II, while we see no effect for SN~Ia relative to SN~II.  We note
some extreme and interesting supernova-host pairs, including the
metal-poor ($\sim 1/4$ solar) host of the recent SN~Ia 2007bk, where the
supernova was found well outside of this dwarf galaxy.  To extend the
luminosity range of supernova hosts to even fainter galaxies, we also
match all the supernovae with $z < 0.3$ to the SDSS-DR6 sky
images, resulting in 1225 matches.  This allows us to identify some even
more extreme cases, such as the recent SN~Ic 2007bg, where the likely
host of this hypernova-like event has an absolute magnitude M$_B \sim
-12$, making it one of the least-luminous supernova hosts ever observed.
This low-luminosity host is certain to be very metal poor ($\sim 1/20$
solar), and therefore this supernova is an excellent candidate for
association with an off-axis GRB.  The two catalogs that we have
constructed are available online and will be updated regularly.
Finally, we discuss various implications of our findings for
understanding supernova progenitors and their host galaxies.
\end{abstract}



\section{Introduction}

On general grounds, it is thought that metallicity will affect the
endpoints of stellar evolution, e.g., the relative outcomes in terms of
different supernova types and the observed properties of each.  Metals
are a source of opacity that affects supernova progenitors
\citep[e.g.,][]{kudritzki00} and also the supernova explosions
themselves \citep[e.g.,][]{heger03}.  However, the hypothesized
metallicity effects have been rather difficult to measure directly.  The
number of supernova progenitors that have been identified directly from
pre-explosion imaging is small and limited to core-collapse events
\citep[e.g.,][]{hendry06, li07}.  Previous works have either used
population studies with only observational proxies for metallicity
\citep[e.g.,][]{prantzos03} or have considered direct metallicity
measurements with only relatively small numbers of events
\citep[e.g.,][]{hamuy00, gallagher05, stanek06, modjaz07}.

A new approach is now possible, which we employ in this paper, that
takes advantage of the large sample of well-observed and typed
supernovae.  Due to a fortuitous match in coverage, many of
these supernovae were in galaxies for which the Sloan Digital Sky Survey
(SDSS) has identified the host galaxies and measured their oxygen
abundances from emission lines in their spectra \citep{tremonti04}.
While these are central metallicities for the host galaxies, and are not
measured for each supernova site, they are much more directly connected
to the latter than proxies like the host luminosity.  To further sharpen
our tests, we compare the metallicity distributions of the host galaxies
of SN~Ib/c and SN~Ia to those of SN~II, which are taken as a control
sample.

The progenitors of core-collapse supernovae (SN~II and Ib/c) are massive
stars, either single or in binaries, with initial main sequence masses
$\ga 8\,\Msun$ \citep[e.g.,][]{heger03}.  The presence of hydrogen in
the spectra of SN~II indicates that the massive envelopes are retained
by the progenitors, of which red supergiants are probably the most
common.  However, SN~Ib/c lack hydrogen (SN~Ib) or both hydrogen and
helium (SN~Ic) in their spectra, and are therefore thought to have
Wolf-Rayet (WR) stars as progenitors (see \citealt{crowther07} for a
review).  The latter originate from the most massive stars, and have had
their outer layers stripped off by strong winds.  Thus SN~Ib/c are
thought to have main sequence masses $\ga 30\,\Msun$, which would make
them $\simeq (8/30)^{1.35} \simeq 20\%$ of all core-collapse supernovae,
assuming a Salpeter slope in the high-mass end of the initial mass
function.

Based on theoretical considerations, the effects of line-driven winds
are expected to introduce a metallicity dependence in the minimum mass
necessary to produce WR stars
\citep[e.g.,][]{heger03,eldridge04,vink05}, which in turn can change the
fractions of core-collapse supernovae that explode as SN~II and
SN~Ib/c. Due to the relative frequencies, SN~Ib/c will be more affected
than SN~II. These metallicity effects on the progenitor winds may
strongly affect the rate at which radioactive $^{26}$Al is expelled into
the interstellar medium before decaying
\citep[e.g.,][]{prantzos04,palacios05}, in which case the decays
contribute to the observed diffuse 1.809 MeV gamma-ray line emission
from the Milky Way \citep[e.g.,][]{diehl06}.  While $^{26}$Al appears to
originate in massive stars, it is not yet known how much comes from the
progenitors or the different core-collapse supernova types
\citep[e.g.,][]{prantzos96,higdon04}. For the most massive stars, GRB
progenitors in the collapsar model (e.g., \citealt{macfadyen99},
\citealt{yoon05}), the interplay between metallicity-dependent mass loss
through winds and rotation may be crucial (e.g.,
\citealt{hirschi05}). In all cases, binary progenitors may be more
complicated \citep[e.g.,][]{eldridge07}.

\citet{prantzos03} used the absolute magnitudes of galaxies as a proxy
for their average metallicities, from the luminosity-metallicity
relationship, and found that the number ratio of SN~Ib/c to SN~II
increases with metallicity; they argued that their result is consistent
with stellar evolution models of massive stars with rotation
\citep[e.g.,][]{meynet06}. If so, then one would expect a more robust
signature if the host metallicities were known directly.  Ideally, in
the latter approach, one would use the metallicities as measured from
follow-up spectra obtained at the supernova sites, but this is difficult
in practice. This approach of using measured as opposed to estimated
metallicities was used by \citet{stanek06} (with compiled results from
the literature) to study nearby long-duration GRBs with subsequent
supernovae, finding that all of them had very low metallicity
environments and that this appeared to be key to forming powerful GRB
jets, and by \citet{modjaz07} to study nearby broad-lined SN~Ic (without
GRBs), finding in contrast that the metallicities of these environments
were much higher. The main caveats associated with these results are the
low statistics, five and twelve events, respectively. We try to combine
the virtues of these two approaches, with higher statistics and mostly
direct metallicity measurements.

The likely progenitors of SN~Ia are white dwarfs, forming from stars
with initial main-sequence masses $\la 8\,\Msun$, which accrete mass
from a companion (single-degenerate model) until they reach the
Chandrasekhar mass ($\simeq 1.4\,\Msun$) and produce a thermonuclear
explosion that completely disrupts the star \citep[e.g.,][]{whelan73}.
During the accretion process, white dwarfs could have strong winds when
the accretion rate reaches a critical value \citep[e.g.,][]{hachisu96},
which would allow them to burn hydrogen steadily and grow in mass. At
low metallicities (${\rm [Fe/H]} \la -1$), SN~Ia may be inhibited
through the single-degenerate channel \citep{kobayashi98}, as the white
dwarf wind is thought to be weak and the system passes through a common
envelope phase before reaching the Chandrasekhar mass. Metallicity also
affects the CNO abundances of white dwarfs, which can affect the
production of $^{56}$Ni in the explosion, and therefore the peak
luminosities of SN~Ia
\citep[e.g.,][]{umeda99,hoeflich00,timmes03,roepke06}. Studies of the
integrated metallicities of nearby SN~Ia hosts
\citep{hamuy00,gallagher05} have shown that metallicity does not seem to
be the main factor regulating their peak luminosities, which is
consistent with some theoretical models
\citep[e.g.,][]{podsiadlowski06}.  Instead, the age of the stellar
population where SN~Ia progenitors originate seems to be very important:
{\it prompt} (SN~Ia explode $\sim 10^{8}$~yr after star formation) and
{\it delayed} (SN~Ia explode $> 10^{9}$~yr after star formation)
components were suggested to explain the high rates of SN~Ia in actively
star-forming galaxies (late type spirals and irregulars) compared with
SN~Ia in old, elliptical galaxies
\citep[e.g.,][]{mannucci05,scannapieco05,neill06}.

In this work, to our knowledge for the first time, we compare the
directly measured oxygen abundances of the hosts of SN~Ib/c and SN~Ia
with SN~II.  We use the Sternberg Astronomical Institute (SAI) supernova
catalog and match it with the SDSS-DR4 catalog of oxygen abundances of a
large sample of star-forming galaxies from SDSS. Using the supernova
classifications presented in the literature, we can separate the sample
according to different supernova types and make statistical comparisons
of the metallicity distributions of their host galaxies.  We also
investigate some individual cases in metal-poor environments that are
especially interesting and which can be used to test the strong
predictions made by some theoretical models.  We create a second catalog
by matching the positions of all supernovae with images from
SDSS-DR6, independent of the host galaxy association.  This allows us to
investigate significantly fainter SNe hosts, and we identify some even
more extreme hosts for follow-up observations. To enable their further
use in other studies, we make both catalogs available online, and will
update them regularly.


\section{First Catalog: Supernova-Host Pairs with Known Host
Metallicities (SAI~$\cap$~SDSS-DR4)}
\label{sec:data1}

We use the SAI supernova catalog\footnote{\scriptsize{{\tt
http://www.sai.msu.su/sn/sncat/}}} \citep{tsvetkov04} to obtain the
main properties of supernovae (name, classification, RA, DEC,
redshift) and their host information when available (galaxy
name, RA, DEC, redshift).  The SAI catalog is a compilation of
information about supernova discoveries, obtained mainly
from reports in the International Astronomical Union Circulars (IAUC),
which include the coordinates and classification of the supernovae
from the IAUCs and also basic information about the host galaxies in
the cases where the galaxies can be identified in online galaxy
catalogs (e.g., HyperLEDA, NED and SDSS). The version of the catalog
we use contains 4,169 entries\footnote{Version updated on June 15,
2007.}, of which we have selected 3,050 supernovae discovered between
1909 and 2007 classified as SN~Ia, II, and Ib/c, including their
sub-types. Supernovae in the catalog with no classification or only
classified as Type~I are not considered for further analysis since we
want to be able to distinguish between SN~Ia and the core-collapse
types SN~Ib/c.
 
\citet{tremonti04} determined metallicities for a sample of star-forming
galaxies in the SDSS Data Release 2 (SDSS-DR2; \citealt{abazajian04})
from their spectra. Here we use a larger sample of 141,317 star-forming
galaxies (excluding AGN) from the SDSS-DR4 \citep{adelman06}, with
metallicities derived in the same consistent fashion, and which are
available online\footnote{\scriptsize{{\tt
http://www.mpa-garching.mpg.de/SDSS/DR4/}}}. The metallicities are
derived by a likelihood analysis which compares multiple nebular
emission lines ([\ion{O}{2}], H$\beta$, [\ion{O}{3}], H$\alpha$,
[\ion{N}{2}], [\ion{S}{2}]) to the predictions of the hybrid
stellar-population plus photoionization models of \citet{charlot01}.  A
particular combination of nebular emission line ratios arises from a
model galaxy that is characterized by a galaxy-averaged metallicity,
ionization parameter, dust-to-metal ratio, and 5500\AA\ dust
attenuation. For each galaxy, a likelihood distribution for metallicity
is constructed by comparison to a large library of model galaxies. We
use the median of the oxygen abundance distributions in this paper. The
metallicities derived by \citet{tremonti04} are essentially on the
\citet{kewley02} abundance scale ($\Delta[\loh] < 0.05$~dex;
\citealt{ellison05}). For further reference in this paper, we call this
galaxy metallicity catalog SDSS-DR4.

We restrict the initial sample of galaxies to 125,958 by applying two of
the cuts that \citet{tremonti04} used for their final cleaned sample:
(1) the redshifts of the galaxies have to be reliable by SDSS standards;
and (2) H$\beta$, H$\alpha$, and [\ion{N}{2}]~$\lambda$6584 should be
detected at $> 5\sigma$ confidence, and
[\ion{S}{2}]~$\lambda\lambda$6717,6731 and [\ion{O}{3}]~$\lambda$5007
should at least have detections.  While in our analysis we directly
compare nebular oxygen abundance within the SDSS-DR4 catalog for the
supernova hosts, when referring to ``Solar metallicity,'' we adopt the
Solar oxygen abundance of $\loh = 8.86$ \citep{delahaye06}.

We cross-matched the SAI catalog with the galaxy metallicity catalog
SDSS-DR4 using a matching radius of 60$\arcsec$ ($\sim 48$~kpc at
$z=0.04$). We used the coordinates of the host galaxies in the cases
where they are known and identified in the SAI catalog, and the
supernovae coordinates were used otherwise.  We also required that the
redshifts reported in the SAI catalog, which were taken from galaxy
catalogs and the IAUCs, to be consistent within $20\%$ with the
redshifts of the closest galaxy from the SDSS catalog that passed the
proximity cut. After selecting the supernovae that passed the proximity
and redshift criteria, we visually inspected the SDSS images around the
galaxies to identify the ones that were wrongly selected as hosts (e.g.,
close galaxy pairs).  The number of supernovae that passed
all these cuts is 254 in total: 95 SN~Ia, 123 SN~II, and 36
SN~Ib/c. There were some galaxies that hosted more than one supernova:
five galaxies had three supernovae each (NGC~1084, NGC~3627, NGC~3631,
NGC~3938, and NGC~5457) and 15 galaxies had two supernovae (NGC~2532,
NGC~2608, NGC~3627, NGC~3780, NGC~3811, NGC~3913, NGC~4012, NGC~4568,
NGC~5584, NGC~5630, NGC~6962, UGC~4132, UGC~5695, IC~4229, and
MCG~+07-34-134).

In Table~\ref{tab1} we present the final matched sample of supernovae
and host galaxy metallicities from SDSS-DR4, as well as the absolute
M$_B$ magnitudes of the galaxies obtained from the HyperLEDA database
and SDSS.  The absolute magnitudes for SDSS galaxies were calculated
using Petrosian $gr$ magnitudes transformed to $B$ magnitudes using the
transformation of \citet{lupton05}, corrected by Galactic extinction
\citep{sfd98} and internal extinction to a face-on geometry
\citep{tully98}, and $k$-corrections \citep{blanton03}. To calculate the
absolute magnitudes, we use a flat cosmology with ${\rm
H}_{0}=70\,\,{\rm km\,s^{-1}\,Mpc^{-1}}$, $\Omega_{M}=0.3$,
$\Omega_{\Lambda}=0.7$. The typical 1$\sigma$ uncertainties in the
oxygen abundances are 0.05~dex at $\loh > 8.5$, and 0.15~dex at $\loh <
8.5$. Our estimated uncertainty in the absolute magnitudes of the hosts
is $\sim 0.3$~mag, calculated from a sub-sample of galaxies in the
catalog with reliable absolute magnitudes from SDSS and HyperLEDA.

Our first catalog, SAI~$\cap$~SDSS-DR4, is available
online\footnote{{\scriptsize{\tt
http://www.astronomy.ohio-state.edu/$\sim$prieto/snhosts/}}} and will be
updated as new supernovae are discovered with host galaxy metallicities
in the SDSS-DR4 catalog. It includes the information presented in
Table~\ref{tab1}, as well as images around the supernovae obtained from
SDSS-DR6.

\begin{figure*}[t]
\epsscale{1.0}
\plotone{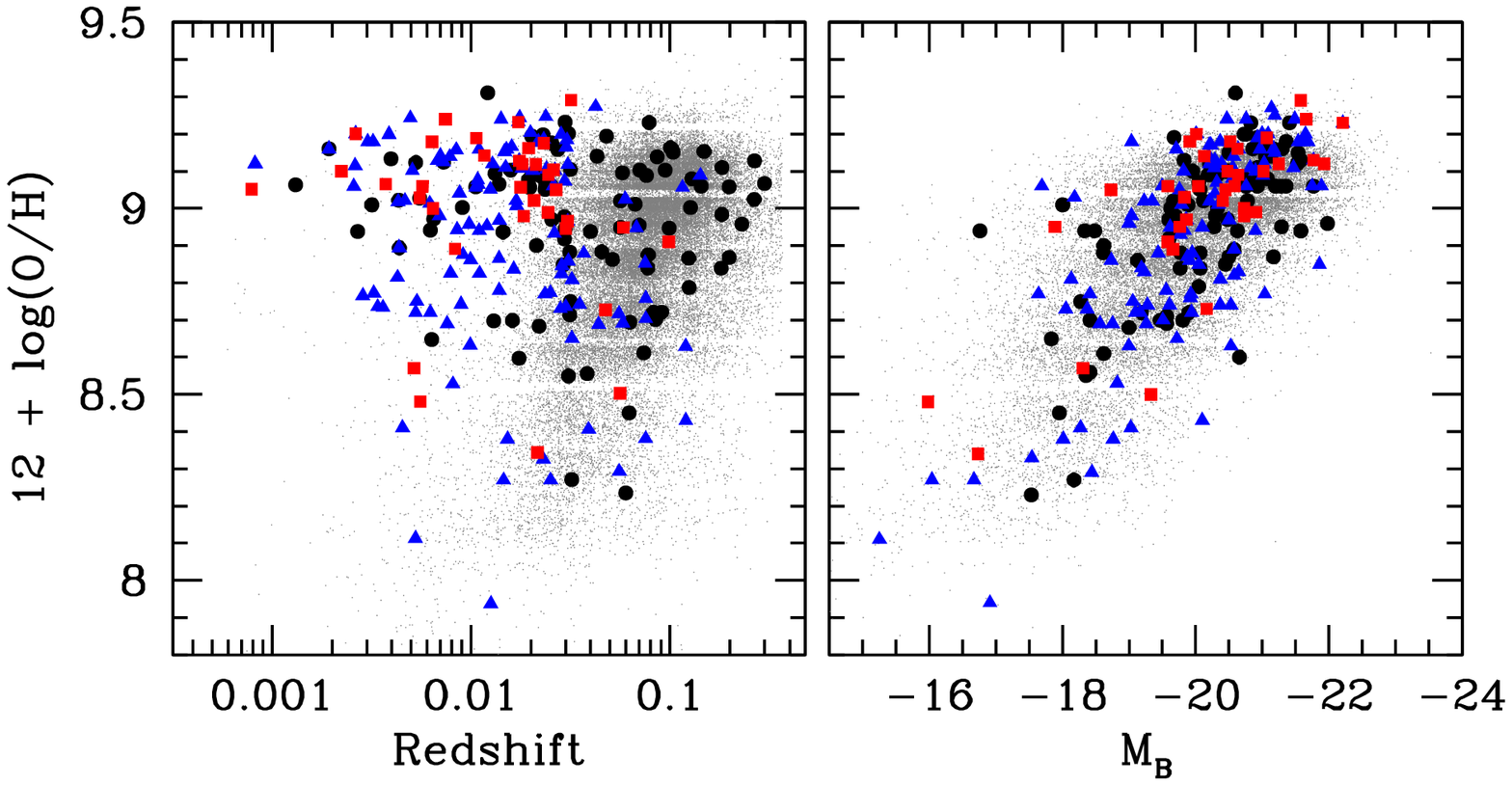}
\caption{Metallicities of supernova host galaxies from SDSS-DR4 as a
function of redshift and absolute $B$ magnitude. The symbols distinguish
different supernova types: SN~II ({\it triangles}), SN~Ib/c ({\it
squares}) and SN~Ia ({\it circles}). The {\it dots} in the {\it left
panel} are 125,958 star forming galaxies in SDSS-DR4 with reliable
metallicity and redshift measurements. The {\it dots} in the {\it right
panel} are a subsample of 86,914 star forming galaxies ($z>0.005$)
selected from the main SDSS-DR4 galaxy sample. \label{fig1} }
\end{figure*}

Figure~\ref{fig1} shows the distribution of metallicities as a function
of redshift and M$_B$ of the supernova host galaxies, as well as the
distribution of star-forming galaxies in the SDSS-DR4 catalog.  The
apparent ``stripes'' in the plots, regions with very few oxygen
abundance measurements, are an effect of the grid of model parameters
(metallicity, ionization parameters, attenuation, etc.) used to
calculate the metallicities (see \citealt{brinchmann04} for details). As
can be seen, the redshift distribution of supernovae varies for
different types, with the median redshifts of the samples at $z=0.014$
(II), 0.018 (Ib/c), and 0.031 (Ia).  This is a combination of several
effects.  First, SN~Ia supernovae are, on average, $\sim 2$~mag brighter
at peak luminosity than core-collapse events \citep{richardson02},
therefore, they can be found at larger distances in magnitude-limited
surveys.  Second, the local rate of core-collapse supernovae in
late-type galaxies is $\sim 3$ times higher than the SN~Ia rate
\citep{cappellaro99,mannucci05}.  Finally, the great interest in SN~Ia
as cosmological distance indicators makes most of the supernovae
searches concentrate their limited spectroscopic follow-up resources on
likely Type~Ia supernovae (as determined by their light curves).

As shown in Figure~\ref{fig1}, the distribution of host galaxy
metallicities follows the distribution of galaxies from SDSS, with a
wide range spanning $\sim 1.4$~dex ($7.9 < \loh < 9.3$). However, there
appear to be significant differences present between the hosts of
different supernovae types. In particular, most of the SN~Ib/c hosts are
concentrated in the higher metallicity/luminosity end of the
distribution ($\loh \ga 8.7$), while the metallicities of SN~II and
SN~Ia hosts are more evenly distributed and appear to be tracing each
other fairly well.

\begin{figure*}[t]
\epsscale{0.8}
\plotone{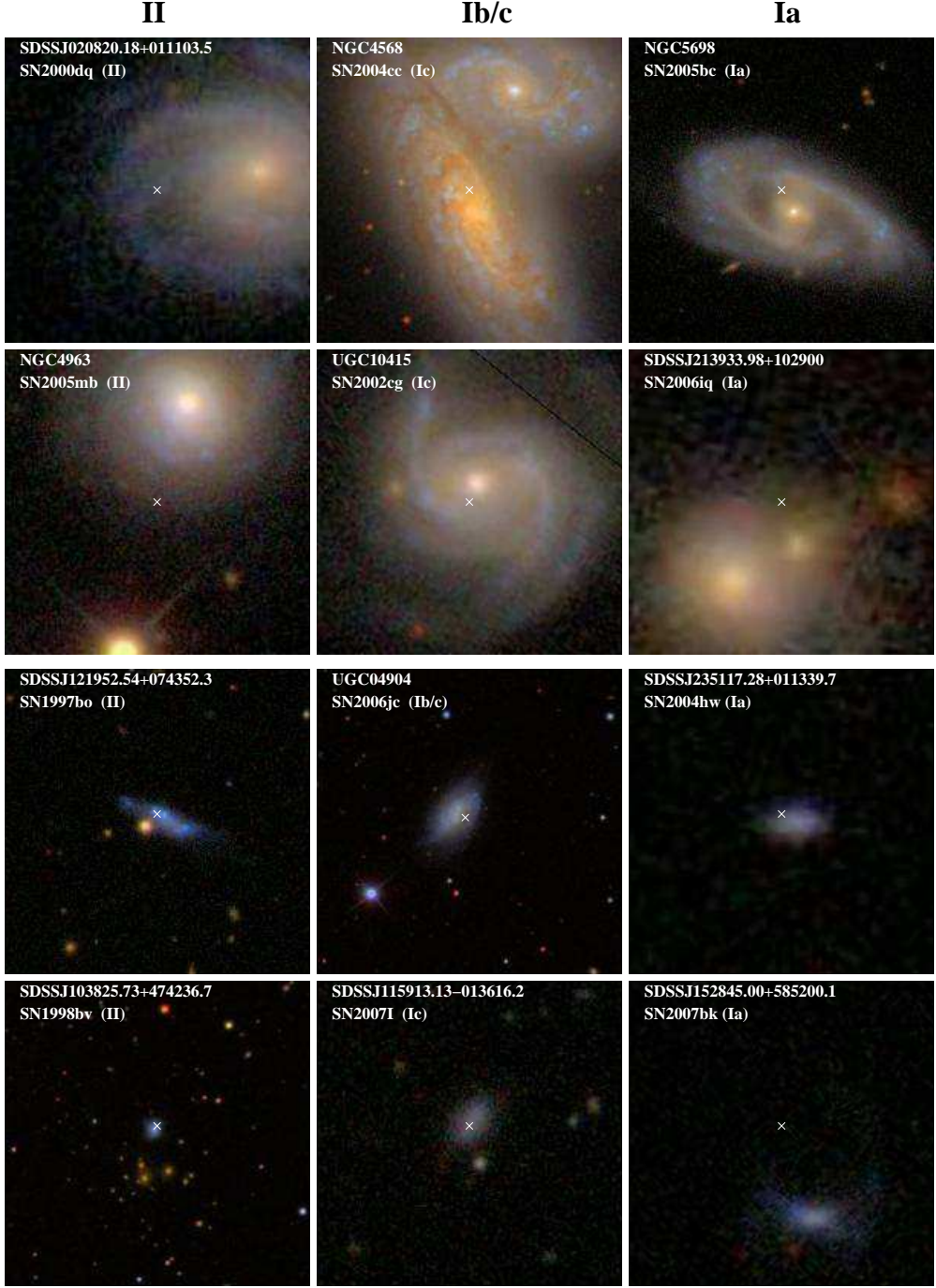}
\caption{SDSS color images of the most metal rich ({\it top six images})
and most metal poor ({\it bottom six images}) host galaxies in our
sample. We show two galaxies of each supernova type. The images are
centered on the position of the supernova explosion, marked with a
cross, with North-up and East-left. They have the same physical size of
30~kpc at the distance of each galaxy. \label{fig2}}
\end{figure*}

Figure~\ref{fig2} shows a mosaic of SDSS-DR6 \citep{adelman07}
images\footnote{{\scriptsize{{\tt
http://cas.sdss.org/dr6/en/tools/chart/chart.asp}}}} of the host
galaxies with the highest and lowest metallicities in the sample,
including two supernovae of each type. This figure shows the wide range
of host galaxy environments present in the sample, from big spirals
(e.g., SN~2000dq, SN~2004cc, SN~2005bc, SN~2005mb, SN~2002cg, and
SN~2006iq) to small dwarfs (e.g., SN~1997bo, SN~2006jc, SN~2004hw,
SN~1998bv, SN~2007I, and SN~2007bk), and that all types of supernovae
can be found in metal-rich and metal-poor star-forming galaxies.


\subsection{Testing Supernova Trends with Metallicity}
\label{sec:analysis}

Is the tendency of SN~Ib/c hosts towards higher metallicity, compared
with SN~II and SN~Ia, clearly seen in Figure~\ref{fig1}, a real physical
effect?  To answer this question we identify and try to reduce some of
the biases present in the sample.

The supernova sample studied in this work is far from homogeneous. The
supernovae have been discovered by a variety of supernova surveys,
including amateur searches that look repeatedly at bright galaxies in
the local universe, professional searches that look at a number of
cataloged galaxies to a certain magnitude limit (e.g., LOSS), and
professional searches that look at all the galaxies in a given volume
(e.g., SDSS-II, The Supernova Factory), among others.  The host galaxies
of supernovae discovered by amateur searches tend to have higher
metallicities due to the luminosity-metallicity relation (see
Figure~\ref{fig1}), while the metallicities of galaxies observed by
professional searches span a wider range.

As an example of a possible bias in the supernovae in our catalog, we
note that the median metallicity decreases by $\sim 0.1$~dex for the
hosts of supernovae discovered between 1970 and 2007. Ideally, all the
supernovae for the current study would be selected from galaxy-impartial
surveys. However, the numbers of different supernova types found by such
surveys in our catalog are still small (especially core-collapse
events), and do not allow a statistical comparison (see the discussion
in \citealt{modjaz07}).

Another bias present in the galaxy data that we use is the so-called
aperture bias \citep{kewley05,ellison05}. The SDSS spectra are taken
with a fixed fiber aperture of $3\arcsec$ (2.4~kpc at $z=0.04$). Since
galaxies have radial metallicity gradients (e.g., \citealt{zaritsky94}),
for nearby galaxies we are, on average, only measuring the higher
central metallicity, while for more distant galaxies we are covering a
larger fraction of the galaxy light with the SDSS fiber. This effect
also depends on galaxy luminosity, as for dwarf galaxies the fiber will
cover a larger fraction of the total light than in large
spirals. \citet{kewley05} find a mean difference of $\sim 0.1$~dex,
although with a large scatter ($0.1-0.3$~dex), between the central and
integrated metallicities of a sample of $\sim 100$ galaxies of all types
(S0-Im) in the redshift range $z = 0.005-0.014$.

In order to reduce these and other biases, we limit the comparison of
supernova types to host galaxies in the redshift range $0.01 < z <
0.04$, where there are 115 supernovae.  In this {\it pseudo}
volume-limited sample, the median redshifts of the 58 SN~II (0.020), 19
SN~Ib/c (0.021) and 38 SN~Ia (0.024) hosts are consistent, while the
number of galaxies in each sub-sample still allows us to make a
meaningful statistical comparison.  By using a small redshift slice we
are, effectively, reducing the aperture biases when comparing the galaxy
metallicity measurements, such that they are now comparable to or
smaller than the statistical errors in the metallicity determination.

We made additional checks of relative biases between supernova types in
our redshift-limited sample.  First, the ratios of the numbers of
SN~Ib/c and SN~Ia to the total number of core-collapse supernovae are
reasonably consistent with the ratios obtained from the local supernovae
rates \citep[e.g.,][]{cappellaro99,mannucci05}. Second, the fact that
the SN--host separation distributions for SN~Ia and SN~II agree,
particularly at small radii (see below), suggests that our supernova
samples are not biased (relatively, one supernova type to another) by
obscuration effects.

We compare the metallicity distributions of the hosts of SN~Ib/c and
SN~Ia to SN~II, which are taken as the control sample. Given that SN~II
are the most common type of supernovae \citep[e.g.,][]{mannucci05} and
that they come from massive stars from a wide range of masses that
explode in all environments, presumably independent of metallicity, they
are effectively giving us the star-formation-rate weighting of the
luminosity-metallicity (or mass-metallicity) relationship for
star-forming galaxies. It would be of interest to test if indeed the
SN~II rates are independent of metallicity, but this is outside the
scope of the current paper.

Figure~\ref{fig3} shows the cumulative distribution of metallicities for
hosts of different supernova types in the redshift ranges $z<0.04$ and
$0.01 < z < 0.04$.  Two important results are immediately apparent:

\begin{itemize}
 
\item The metallicities of SN~Ib/c hosts tend to be higher than those
of SN~II hosts,

\item The SN~Ia and SN~II hosts have very similar metallicity
  distributions.

\end{itemize}

Kolmogorov-Smirnov (KS) tests between the metallicity distributions of
different supernova types in the redshift range $0.01 < z < 0.04$
strengthen these findings. The KS probabilities of two host metallicity
samples being drawn from the same distribution are: 5\% (II--Ib/c), 3\%
(Ia--Ib/c) and 56\% (II--Ia). We obtain a similar result if we compare
the mean metallicities of the host samples: 8.94$\pm$0.04 (SN~II),
8.94$\pm$0.04 (SN~Ia) and 9.06$\pm$0.04 (SN~Ib/c), where the errors are
the RMS of similar-sized samples obtained using bootstrap resampling.

The metallicity distribution of the SDSS-DR4 star-forming galaxies in
our redshift range, weighted only by galaxy counts, is also shown in
Figure~\ref{fig3}. This should not be used in any comparisons, as it
does not take into account the weighting with star formation rate or the
supernova and galaxy selection criteria. We take all of these into
account by only making relative comparisons between supernova types.

If we restrict the sample of SN~Ib/c to only SN~Ic and broad-lined Ic in
the same redshift range, leaving out supernovae classified as SN~Ib/c
and SN~Ib, the difference in metallicity distributions of the hosts of
SN~II and SN~Ic$+$hypernovae (13 SN) becomes smaller, with a KS
probability of 19\%.  If only the three supernova classified as SN~Ib
(SN~2003I, SN~2005O, and SN~2005hl) in the {\it pseudo} volume-limited
sample are not considered, then the KS probability of SN~II and
SN~Ic$+$hypernovae$+$SNIb/c being drawn from the same sample is 10\%.

In Figure~\ref{fig4} we show the number ratio of SN~Ib/c to SN~II as a
function of the metallicities of the host galaxies. This ratio is very
important because the rates of core-collapse SNe are expected to change
as a function of the progenitor mass and metallicity and, therefore, it
can help to put constrains on massive stellar evolution models
\citep[e.g.,][]{eldridge07}. We have calculated the ratio in bins of
equal number of SN~II+SN~Ib/c, with 11 SNe per bin, to do a direct
comparison with the results of \citet{prantzos03}. The small statistics
compared with \citet{prantzos03}, who used the absolute magnitudes of
the hosts as a proxy for the average metallities through the
luminosity-metallicity relationship, is reflected in the large errors of
the ratio. The large error bars do not allow us to put constraints in
progenitor models, however, the general trend observed in the cumulative
distribution (see Figure~\ref{fig3}) is confirmed with the number
counts: SN~Ib/c are more common at higher metallicities compared with
SN~II. Our results are consistent with those of \citet{prantzos03}.

Figure~\ref{fig5} shows the cumulative distributions of projected
host-supernova distances for the reduced sample of 115 SNe used to
compare the host metallicities ($0.01 < z < 0.04$). Clearly, the SN~Ib/c
in the sample are found more towards the centers of their hosts when
compared with SN~II and SN~Ia
\citep[e.g.,][]{vandenbergh97,tsvetkov04,petrosian05}, which have
similar distributions to each other (as also in Figure~\ref{fig3}).  A
galactocentric concentration of SN~Ib/c and their progenitors may be
important for the angular distributions of diffuse gamma-ray line
emission from the Milky Way. Besides the 1.809 MeV line from $^{26}$Al,
the 0.511 MeV line from positron annihilation is poorly understood, in
terms of its high flux and very strong central concentration
\citep[e.g.,][]{casse04, knodlseder05,beacom06}. Since the SN~Ib/c are
found at small separation, the central galaxy metallicities determined
by the SDSS should be representative of the local environments of the
supernovae.  Taking into account the existence of negative metallicity
gradients in increasing galactocentric radii, the local metallicities of
the SN~II and SN~Ia, if anything, are even {\em lower} than deduced from
the SDSS central metallicities.  The tendency for SN~Ib/c to prefer
higher metallicity relative to SN~II and SN~Ia is probably even stronger
than shown in Figure~\ref{fig3}.


\subsection{Supernovae in Low-Metallicity Hosts}

Even though we have shown that there is a strong preference of SN~Ib/c
for high-metallicity environments, compared with SN~II and SN~Ia, there
are four SN~Ib/c with relatively metal-poor host galaxies ($\loh <
8.6$).  These events, and also some SN~Ia found in low-metallicity
dwarfs, made us investigate more carefully a number of individual cases.
We found that among the lowest-metallicity host galaxies in the sample,
there were supernovae that stood out because of their unusual properties
(all shown in Figure~\ref{fig2}).

\begin{description}
  
\item [SN~2006jc:] Peculiar SN~Ib/c supernova with strong \ion{He}{1}
  lines in emission in the spectrum \citep{crotts06}, thought to arise
  from the interaction of the supernova ejecta with a He-rich
  circumstellar medium \citep{foley07,pastorello07}. Its host galaxy,
  UGC~4904 at $z=0.006$, is a low luminosity, blue, and relatively
  low-metallicity starburst (M$_{B}=-16.0$, $\loh=8.5$).  Interestingly,
  the host galaxy of SN~2002ao at $z=0.005$ (in UGC~9299), another
  peculiar SN~Ib/c with spectral properties very similar to SN~2006jc
  \citep{benetti06} that is also present in our first catalog, has low
  metallicity compared with the majority of the SN~Ib/c hosts, and
  shares similar morphological properties with the host of SN~2006jc.
 
\item [SN~2007I:] Broad-lined Ic (or hypernova) with a spectrum similar
  to SN~1997ef \citep{blondin07} at $z=0.022$. Its host galaxy is a
  star-forming, low-metallicity dwarf (M$_{B}=-16.7$, $\loh=8.3$),
  unlike other broad-lined Ic supernovae observed in higher-metallicity
  galaxies \citep{modjaz07}, and somewhat similar to the host galaxies
  of long GRBs associated with supernovae \citep{stanek06,fruchter06};
  however, see a detailed discussion in \citet{modjaz07}.  The other
  four broad-lined Ic supernovae in our sample that have been reported
  in the literature are: SN~1997ef \citep{iwamoto00}, 2004ib
  \citep{sako05}, 2005ks \citep{frieman05}, 2006qk \citep{frieman06}.
  
\item [SN~2007bk:] Type~Ia supernova with a spectrum similar to the slow
  decliner/luminous SN~1991T \citep{dennefeld07} at $z=0.032$.  The host
  galaxy is a low metallicity/luminosity dwarf, with M$_B=-18.2$ and
  $\loh=8.3$, similar to the Large Magellanic Cloud.  The supernova was
  found very far from the center of its dwarf host, at a projected
  separation of $\sim9$~kpc. The magnitude of the supernova at discovery
  ($R=16.7$, \citealt{mikuz07}) and the phase obtained from the spectrum
  ($+50$~days, \citealt{dennefeld07}, although S.~Blondin finds equally
  good matches with templates at $+30$~days, private communication),
  imply that this was a very luminous Type~Ia event. If the reported
  discovery magnitude and spectral phases are accurate, SN~2007bk was
  $\sim 0.5-1.5$~mag brighter than SN~1991T at the same phase after
  maximum light.

\end{description}


\section{Second Catalog: Supernova-Host Pairs with Unknown Host
Metallicities (SAI~$\cap$~SDSS-DR6)}
\label{sec:data2}

The existence of supernovae with unusual properties among the most
metal-poor, low-luminosity galaxies in the first catalog prompted us to
investigate a much larger sample of supernovae. We constructed a second
catalog with images around the positions of supernovae using SDSS,
matching the SAI catalog with SDSS-DR6.  We included the redshifts
obtained from the SAI catalog to produce images in physical units around
the supernovae. The total number of matches is 1225 for supernovae at
$z<0.3$. This catalog is also available online, with the first catalog
described earlier in \S~\ref{sec:data1}.

This extended second catalog (SAI~$\cap$~SDSS-DR6) does not have
information about metallicities or luminosities of the hosts. It is a
visual tool that can be used to explore the environments around
supernovae found in the SDSS area, independent of the host galaxy
association. Identification of the supernovae hosts and their integrated
properties obtained from SDSS will be added in a future study.

Visually inspecting the images of the second catalog, we identified a
number of supernovae in what appear to be very faint galaxies, and which
are likely to be low-luminosity, metal-poor galaxies not present in the
first catalog. Some examples in the redshift range $0.01 \la z \la 0.05$
are (supernova types shown in parentheses): SN~1997ab (II), SN~1997az
(II), SN~2001bk (II), SN~2003cv (II), SN~2004gy (II), SN~2004hx (II),
SN~2005cg (Ia), SN~2005gi (II), SN~2005hm (Ib), SN~2005lb (II), SN~2006L
(IIn), SN~2006bx (II), SN~2006fg (II), 2007bg (Ic), 2007bu (II), and
2007ce (Ic). In this incomplete sample, which was selected by noting
some especially low-luminosity galaxies, the SN~Ia/SN~II ratio is lower
than expected.  Similarly, in our catalog of hosts with known
metallicities, SN~Ia may also be relatively underabundant at the lowest
host luminosities and metallicities, as shown in Figure~\ref{fig1}. We
caution that the small statistics make these only hints, and we discuss
these issues further below.

One of the most interesting supernovae in our second catalog is
SN~2007bg, a recently discovered broad-lined Ic
\citep{quimby07,harutyunyan07} at $z=0.03$, which has an extremely faint
galaxy coincident with the position of the supernova. Using photometry
and images from SDSS-DR6, we estimate the luminosity of the apparent
host galaxy to be M${_B} \approx -12$, most likely a very metal-poor
dwarf ($\loh \sim 7.5$, or $\sim 1/20$~solar; see the
metallicity-luminosity relationship extended to dwarf galaxies by
\citealt{lee06}). Due to the extremely low luminosity of that galaxy, in
fact one of the lowest-luminosity supernova hosts ever seen, and also
fainter than most if not all GRB hosts \citep[see e.g.,][]{fruchter06},
this event may represent the missing link between broad-lined SN~Ic and
GRBs.  This event is therefore an excellent candidate for a search for
an off-axis GRB jet in radio \citep{soderberg06} and possibly other
wavelengths.  Follow-up spectroscopic observations and deep photometry
to determine the metallicity of the host and study the supernova
environment are strongly encouraged in this and other cases of very
low-luminosity SN hosts.


\section{Discussion and Conclusions}
\label{sec:discussion}

\begin{figure*}[t]
\epsscale{1.0}
\plotone{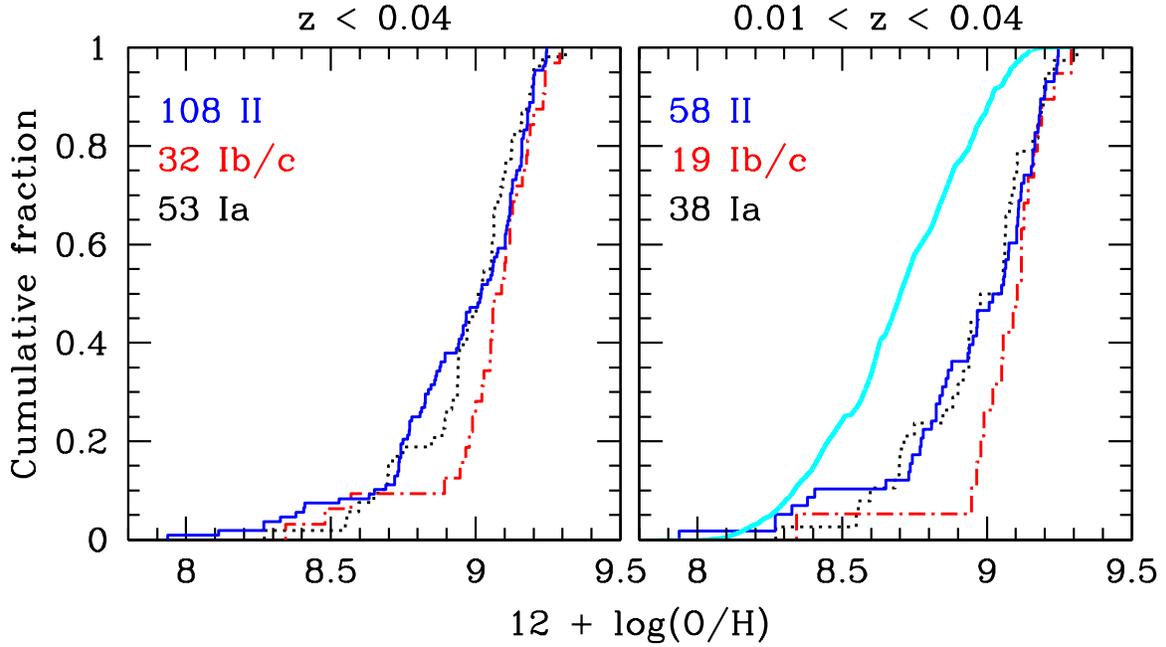}
\caption{Cumulative fraction plots with the oxygen abundance of the
supernova host galaxies. The number of host galaxies of each supernova
type is indicated in each panel and the lines correspond to: SN~II ({\it
solid}), SN~Ib/c ({\it dot-dashed}) and SN~Ia ({\it dotted}). The {\it
left panel} includes host galaxies with redshifts $z < 0.04$, while the
{\it right panel} includes host galaxies with redshifts $0.01 < z <
0.04$. The {\it thick line} in the {\it right panel} shows the
cumulative distribution of the 15,116 star-forming galaxies from
SDSS-DR4 in the same redshift range. \label{fig3}}
\end{figure*}

\begin{figure*}[t]
\epsscale{1.0}
\plotone{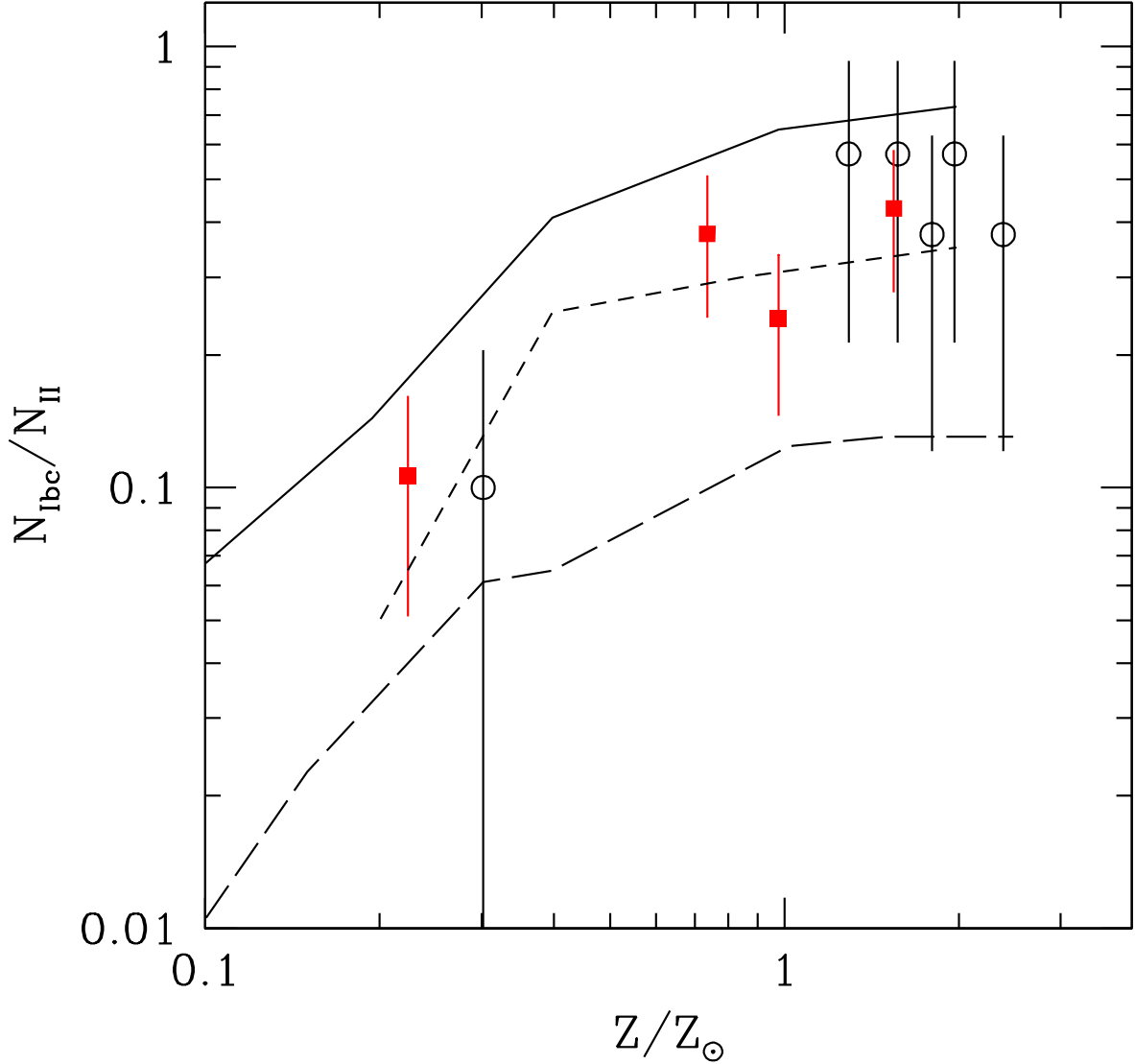}
\caption{Number ratio of SN~Ib/c to SN~II as a function of metallicity
of the host galaxies. The {\em open circles} are the values obtained
with our sample from directly measured central metallicities, and the
{\em filled squares} are the results from \citet{prantzos03} using
absolute magnitudes as a proxy to host metallicities. The error bars are
obtained from Poisson statistics. The {\em solid line} shows the
predicted ratio from the binary models of \citet{eldridge07}; the {\em
dashed line} is from the models of single stars with rotation of
\citet{maeder04}; and the {\em dot-dashed line} is from the single star
models of \citet{eldridge07}. \label{fig4}}
\end{figure*}

\begin{figure*}[t]
\epsscale{1.0} \plotone{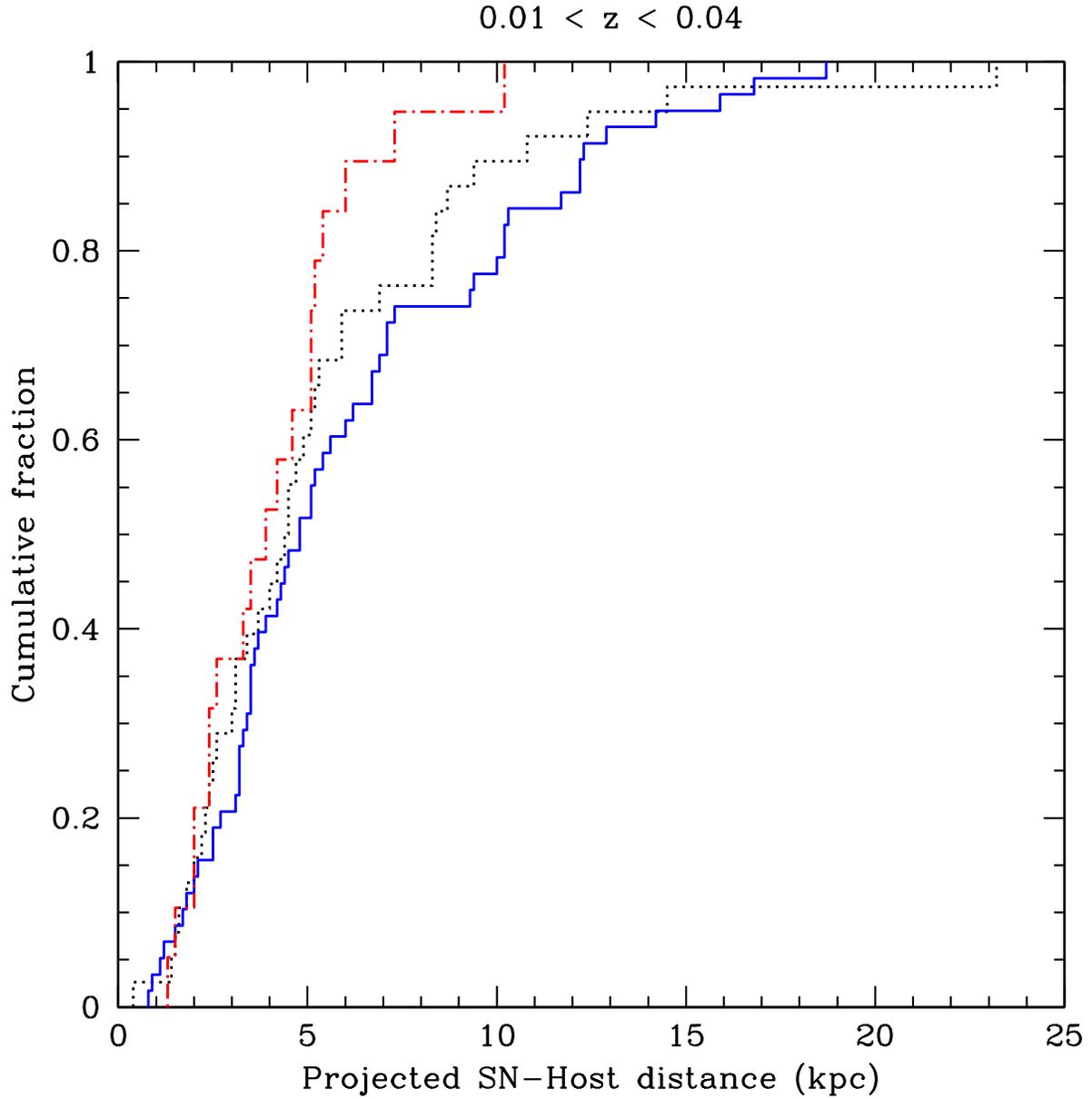}
\caption{Cumulative fraction plot of the projected separation between
the supernova and its host for the reduced sample in the redshift range
$0.01 < z < 0.04$. The lines correspond to: SN~II ({\it solid}), SN~Ib/c
({\it dot-dashed}) and SN~Ia ({\it dotted}). \label{fig5}}
\end{figure*}

We find that SN~Ib/c tend to be in high-metallicity host galaxies,
compared to SN~II, our control sample that traces the underlying star
formation rates. This is the first time that such a trend has been found
using the directly-measured oxygen abundances of the supernova host
galaxies.  This confirms and greatly strengthens an earlier result of
\citet{prantzos03}, which found a similar result using the absolute
magnitudes of the host galaxies as an indirect estimate of their
metallicities through the luminosity-metallicity relationship.  This can
be interpreted in relative supernova rates: the ratio of SN~Ib/c to
SN~II increases with increasing metallicity and hence also cosmic age.
We also find that SN~Ib/c are consistently found towards the centers of
their hosts compared with SN~II and SN~Ia, which had been also found in
previous studies \citep[e.g.,][]{vandenbergh97,tsvetkov04,petrosian05}.
This suggests that direct measurements of metallicities at the explosion
sites, as opposed to the central host metallicities used here, would
reveal an even stronger effect, due to the radial metallicity gradients
observed in spiral galaxies.  The local metallicities of SN~Ib/c would
be less reduced with respect to the central metallicities than SN~II and
SN~Ia, which would widen the separation seen in Figure~\ref{fig3}.

The tendency towards high metallicity of SN~Ib/c environments compared
to those of SN~II supports, in general terms, theoretical models of the
effects of metallicity in stellar evolution and the massive stars that
are core-collapse supernova progenitors (e.g.,
\citealt{heger03,meynet06,eldridge07,fryer07}).  Also, models of stellar
evolution that include rotation, from \citet{meynet06}, predict that at
high metallicity Wolf-Rayet stars will earlier enter the WC phase when
they still are rich in helium, and that these stars would explode as
SN~Ib.  The fact that we do see both SN~Ib and SN~Ic in hosts at high
metallicity should not be taken as inconsistent with these models,
mainly because the number of supernovae is small and the sample has not
been homogeneously selected. There is an indication, although not
statistically significant, that SN~Ib may be more common in higher
metallicity environments than SN~Ic and broad-lined SN~Ic in our sample.

The agreement between the metallicity distributions of the hosts of
SN~II and SN~Ia shows that their hosts are sampling a wide range of
properties of star-forming galaxies, from the relatively metal-poor
dwarfs to metal-rich grand design spirals. Using models of white dwarf
winds in the framework of single-degenerate progenitors of SN~Ia
\citep{hachisu96}, \citet{kobayashi98} made a strong prediction that
SN~Ia would not be found in low metallicity environments, such as dwarf
galaxies and the outskirts of spiral galaxies.  However, we do observe
SN~Ia in metal poor dwarfs (e.g., SN~2004hw, SN~2006oa, and SN~2007bk,
with host metallicities between $\sim 0.2$ and 0.5 solar) and at large
projected distances ($> 10$~kpc) from their star-forming hosts (e.g.,
SN~1988Y, SN~1992B, SN~1993I, SN~2001bg, SN~2002gf, SN~2004ia,
SN~2004ig, SN~2005ms, SN~2006fi, and SN~2006gl). There are also extreme
cases that have been pointed out in previous studies, like the
low-luminosity dwarf (M$_B \approx -12.2$) host galaxy of the luminous
and slow decliner SN~1999aw \citep{strolger02}, which is most likely
very metal-poor ($\loh \sim 7.5$, or $\sim 1/20$~solar; see
\citealt{lee06}).  Also, SN~2005cg was found in a dwarf with subsolar
gas metallicity \citep{quimby06}.

We do not find a statistically significant low-metallicity threshold in
the metallicities of SN~Ia compared with SN~II hosts, as predicted from
theory by \citet{kobayashi98} for single-degenerate progenitors of SN~Ia
with winds. However, there is a preference for finding more SN~II in
very faint galaxies compared with SN~Ia in our second catalog, which is
suggestive of a luminosity or metallicity threshold for the main channel
that produces SN~Ia. This will have to be explored in the future with a
larger sample that includes good luminosity information for the hosts
and actual metallicities measured from spectra. If no metallicity
threshold is found in larger samples, it means that the models and
predictions of white dwarf winds will have to be revisited. This would
have implications for modeling and understanding of galactic chemical
evolution that include the effects of white dwarf winds to shut down
SN~Ia at low metallicities (e.g., \citealt{kobayashi07}). Interestingly,
modeling the X-ray spectra of supernova remnants from probable SN~Ia
explosions in our Galaxy, the LMC and M31, \citet{badenes07} did not
find the strong effects of white dwarf winds predicted from theory.

On the other hand, independent of the existence (or not) of a mechanism
that can shut down the production of SN~Ia in low-metallicity
environments, we have noted examples of SN~Ia that explode in
low-metallicity dwarf galaxies, like SN~2007bk. Also, supernova remnants
from probable SN~Ia have been identified in the LMC
\citep[e.g.,][]{hughes95} and SMC \citep[e.g.,][]{vanderheyden04}. Is
this SN~Ia population dominated by a different kind of progenitors, like
double-degenerate mergers, compared to the main progenitor channel? Is
the expected trend between progenitor metallicities and peak-luminosity
starting to appear as we extend the sample to even lower metallicity
hosts?  It is suggestive that the small number of SN~Ia in
low-metallicity hosts, like SN~2007bk, SN~2005cg and SN~1999aw, were all
luminous events compared with normal SN~Ia. Also, the very luminous
SN~Ia events that have spectral signatures of a strong ejecta-CSM
interaction, like SN~2005gj, are mostly associated with low-luminosity,
and most likely low-metallicity, hosts \citep{prieto07}. Is low
metallicity necessary to produce this extreme class of SN~Ia? Detailed
comparison studies of the observational properties of supernovae in
these extreme environments are encouraged.

In the course of this work, we have prepared two new catalogs that
should be useful for other studies.  We used the SAI supernova catalog
and the SDSS-DR4 sample of metallicities of star-forming galaxies from
\citet{tremonti04} to produce a catalog of supernovae hosts with
metallicities derived in a consistent fashion. From this first catalog,
we found several interesting core-collapse (e.g., SN~2002ao, SN~2006jc,
and SN~2007I) and SN~Ia events (e.g., SN~2007bk) in low-metallicity
galaxies.  We constructed a second catalog by matching the SAI supernova
catalog with images obtained from SDSS-DR6. The second catalog does not
contain information on host metallicities, but it can be used to
investigate the environments of supernovae independent of the host
association. In that second catalog, we found several examples of
core-collapse supernovae in faint galaxies. One of most interesting
cases is SN~2007bg, a broad-lined SN~Ic in an extremely low-luminosity
and very likely low-metallicity host. These catalogs will allow
researchers to select interesting candidates for further follow-up
observations. Also, as more homogeneous light curve and spectroscopic
data become available for supernovae in the first catalog, this will
allow us to test possible correlations between supernova properties and
the metallicities of their hosts, which may turn out to be crucial for
improving our understanding of the nature of different supernova
explosions. Another possible use of our catalogs is for systematically
characterizing the morphologies of supernova hosts.
 
We stress the great importance of galaxy-impartial surveys for finding
and studying the properties of all supernovae types.  Some very
interesting and potentially informative supernovae have been found in
very low-luminosity, low-metallicity galaxies, hosts which are not
included in supernova surveys based on catalogs of normal galaxies.
These unusual supernovae and hosts may help probe the relationship
between the SN~Ib/c and SN~II core-collapse supernova types, the
progenitors of SN~Ia as well as the possible correlations between
observed SN~Ia properties and host metallicities, the supernova-GRB
connection \citep[e.g.,][]{stanek03} and its possible metallicity
dependence \citep[e.g.,][]{stanek06,modjaz07}, and also to test the
consistency between the cosmic stellar birth and death rates
\citep[e.g.,][]{hopkins06}. As we pointed out in \S~\ref{sec:analysis},
presently the comparison of host metallicities using supernovae
discovered by galaxy-impartial surveys is limited by their small
numbers, especially for core-collapse events, since SN~Ia receive much
more attention when decisions about spectroscopic follow-up are made.
This is also true for the study of their observational properties (e.g.,
light curves and spectra).  However, in order to better understand all
types of cosmic explosions and put further constraints on the
predictions of stellar evolution theory, a larger effort on other
supernovae types is greatly needed.

\vspace{0.5in}

We are grateful to C.~Tremonti for help with the extended SDSS-DR4
catalog of galaxy metallicities. We thank C.~Badenes, S.~Blondin,
A.~Hopkins, J.~Johnson, M.~Kistler, M.~Modjaz, and G.~Pojmanski for
helpful discussions and suggestions. JFB is supported by the National
Science Foundation CAREER Grant PHY-0547102.

Funding for the SDSS and SDSS-II has been provided by the Alfred P.
Sloan Foundation, the Participating Institutions, the National Science
Foundation, the U.S. Department of Energy, the National Aeronautics and
Space Administration, the Japanese Monbukagakusho, the Max Planck
Society, and the Higher Education Funding Council for England.

The SDSS is managed by the Astrophysical Research Consortium for the
Participating Institutions. The Participating Institutions are the
American Museum of Natural History, Astrophysical Institute Potsdam,
University of Basel, Cambridge University, Case Western Reserve
University, University of Chicago, Drexel University, Fermilab, the
Institute for Advanced Study, the Japan Participation Group, Johns
Hopkins University, the Joint Institute for Nuclear Astrophysics, the
Kavli Institute for Particle Astrophysics and Cosmology, the Korean
Scientist Group, the Chinese Academy of Sciences (LAMOST), Los Alamos
National Laboratory, the Max-Planck-Institute for Astronomy (MPIA), the
Max-Planck-Institute for Astrophysics (MPA), New Mexico State
University, Ohio State University, University of Pittsburgh, University
of Portsmouth, Princeton University, the United States Naval
Observatory, and the University of Washington.

\newpage

\begin{deluxetable}{lccccccrrccc}
\rotate
\tabletypesize{\scriptsize}
\tablewidth{-1pt}
\tablecaption{Supernova and host galaxy data \label{tab1}}
\tablehead{
\colhead{SN~Name} &
\colhead{Type\tablenotemark{a}} &
\colhead{RA (J2000)} &
\colhead{DEC (J2000)} &
\colhead{Host~Name} &
\colhead{RA (J2000)} &
\colhead{DEC (J2000)} &
\colhead{Distance\tablenotemark{b}} &
\colhead{Distance\tablenotemark{b}} &
\colhead{z\tablenotemark{c}} &
\colhead{M$_{B}$} &
\colhead{12 + log(O/H)} \\
\colhead{} &
\colhead{} &
\colhead{(deg)} &
\colhead{(deg)} &
\colhead{} &
\colhead{(deg)} &
\colhead{(deg)} &
\colhead{(arcsec)} &
\colhead{(kpc)} &
\colhead{} &
\colhead{(mag)} &
\colhead{(dex)}
}
\startdata
     1909A &        II: &   210.5129 &    54.4661 &                        NGC5457 &   210.8022 &    54.3489  &   738.5  & $\ldots$  &    0.00082 &    --21.03 &  9.12  \\ 
     1920A &        II: &   128.8156 &    28.4754 &                        NGC2608 &   128.8222 &    28.4734  &    22.0  &    3.6  &    0.00727 &     --20.21 &  9.12  \\ 
     1936A &        IIL &   184.9830 &     5.3522 &                        NGC4273 &   184.9836 &     5.3433  &    32.2  &    6.0  &    0.00783 &     --20.57 &  9.14  \\  
     ...       &    ...        &    ...        &   ...         &   ...            &  ...          &    ...        &   ...         & ...           & ... \\
    ...        &    ...        &    ...        &   ...         &   ...            &  ...          &    ...        &   ...         & ...           & ... \\
    2007av &         II &   158.6798 &    11.1938 &                        NGC3279 &   158.6783 &    11.1974  &    13.8  & $\ldots$  &    0.00464 &    --19.35 &  9.02  \\ 
    2007be &       IIP: &   189.5277 &   --0.0309 &                       UGC07800 &   189.5223 &   --0.0270  &    24.1  &    6.7  &    0.01251 &     --19.76 &  9.05  \\ 
    2007bk &         Ia &   232.1899 &    58.8702 &        SDSSJ152845.00+585200.1 &   232.1875 &    58.8667  &    13.4  &    8.7  &    0.03214 &     --18.17 &  8.27  \\  
\enddata     
\tablecomments{Complete table is available online.}
\tablenotetext{a}{Supernova classification in the SAI Catalog. The types followed by a colon indicate a provisional classification in the SAI Catalog.}
\tablenotetext{b}{Projected SN-Host distance.}
\tablenotetext{c}{Redshift of the host galaxy from SDSS-DR4.}
\end{deluxetable}


\newpage

\begin{thebibliography}

\bibitem[Abazajian et~al.(2004)]{abazajian04} Abazajian, K., et~al. 
2004, \aj, 128, 502  

\bibitem[Adelman-McCarthy et~al.(2006)]{adelman06} Adelman-McCarthy, J.~K., et~al.
2006, \apjs, 162, 38

\bibitem[Adelman-McCarthy et~al.(2007)]{adelman07} Adelman-McCarthy, J.~K., et~al.
2007, \apjs, submitted  (arxiv:0707.3413)

\bibitem[Badenes et~al.(2007)]{badenes07} Badenes, C., et~al. 2007,
\apj, 662, 472

\bibitem[Beacom \& Y{\"u}ksel(2006)]{beacom06} Beacom, J.~F., \&
Y{\"u}ksel, H. 2006, \prl, 97, 071102

\bibitem[Benetti et~al.(2006)]{benetti06} Benetti, S., et~al. 2006, CBET~674

\bibitem[Blanton et~al.(2003)]{blanton03} Blanton, M.~R., et~al.
2003, \aj, 125, 2348

\bibitem[Blondin et~al.(2007)]{blondin07} Blondin, S., et~al. 2007, 
CBET~808

\bibitem[Brinchmann et~al.(2004)]{brinchmann04} Brinchmann, J.,
et~al. 2004, \mnras, 351, 1151

\bibitem[Cappellaro et~al.(1999)]{cappellaro99} Cappellaro, E., et~al. 1999,
\aap, 351, 459

\bibitem[Cass\'e et~al.(2004)]{casse04} Cass\'e, M., et~al. 2004, \apj,
602, 17

\bibitem[Charlot \& Longhetti(2001)]{charlot01} Charlot, S., \&
Longhetti, M. 2001, \mnras, 323, 887

\bibitem[Crotts et~al.(2006)]{crotts06} Crotts, A., et~al. 2006, CBET~672 

\bibitem[Crowther(2007)]{crowther07} Crowther, P.~A. 2007, \araa, 45,
177

\bibitem[Delahaye \& Pinsonneault(2006)]{delahaye06} Delahaye, F., \&
Pinsonneault, M.~H. 2006, \apj, 649, 529

\bibitem[Dennefeld et~al.(2007)]{dennefeld07} Dennefeld, M., et~al. 2007, CBET~937

\bibitem[Diehl et~al.(2006)]{diehl06} Diehl, R., et~al. 2006, Nature,
439, 45

\bibitem[Eldridge \& Tout(2004)]{eldridge04} Eldridge, J.~J., \& Tout,
C.~A. 2004, \mnras, 353, 87

\bibitem[Eldridge(2007)]{eldridge07} Eldridge, J.~J. 2007, \rmxaa, 30, 35

\bibitem[Ellison \& Kewley(2005)]{ellison05} Ellison, S.~L., \& Kewley,
L.~J., astro-ph/0508627

\bibitem[Foley et~al.(2007)]{foley07} Foley R.~J., et~al. 2007, \apjl, 657, L105  

\bibitem[Frieman et~al.(2005)]{frieman05} Frieman, J., et~al. 2005, IAUC~8616

\bibitem[Frieman et~al.(2006)]{frieman06} Frieman, J., et~al. 2006, CBET~762

\bibitem[Fruchter et~al.(2006)]{fruchter06} Fruchter, A.~S.,
et~al. 2006, Nature, 441, 463

\bibitem[Fryer et~al.(2007)]{fryer07} Fryer, C., et~al. 2007, astro-ph/0702338

\bibitem[Gallagher et~al.(2005)]{gallagher05} Gallagher, J.~S., et~al. 2005, \apj, 634, 210

\bibitem[Hachisu et~al.(1996)]{hachisu96} Hachisu, I., Kato, M., \& Nomoto, K. 1996, \apjl, 470, L97 

\bibitem[Hamuy et~al.(2000)]{hamuy00} Hamuy, M., et~al. 2000, \aj, 120, 1479 

\bibitem[Harutyunyan et~al.(2007)]{harutyunyan07} Harutyunyan, A.,
et~al. 2007, CBET~948

\bibitem[Heger et~al.(2003)]{heger03} Heger, A., et~al. 2003, \apj, 591, 288

\bibitem[Hendry et~al.(2006)]{hendry06} Hendry, M.~A., et~al. 2006,
\mnras, 369, 1303

\bibitem[Higdon et~al.(2004)]{higdon04} Higdon, J.~C., et~al. 2004,
\apjl, 611, 29L

\bibitem[Hirschi et~al.(2005)]{hirschi05} Hirschi, R., Meynet, G., \&
Maeder, A. 2005, \aap, 443, 581

\bibitem[H{\"o}flich et~al.(2000)]{hoeflich00} H{\"o}flich, P., et~al
2000, \apj, 528, 590

\bibitem[Hopkins \& Beacom(2006)]{hopkins06} Hopkins, A.~M., \& Beacom,
J.~F. 2006, \apj, 651, 142

\bibitem[Hughes et~al.(1995)]{hughes95} Hughes, J.~P., et~al. 1995,
\apjl, 444, L81

\bibitem[Iwamoto et~al.(2000)]{iwamoto00} Iwamoto, K., et~al. 2000,
\apj, 534, 660

\bibitem[Kewley \& Dopita(2002)]{kewley02} Kewley, L.~J., \& Dopita,
M.~A. 2002, \apjs, 142, 35

\bibitem[Kewley et~al.(2005)]{kewley05} Kewley, L.~J., Jansen, R.~A., \& Geller, M.~J.
2005, \pasp, 117, 227

\bibitem[Kn{\"o}dlseder et~al.(2005)]{knodlseder05} Kn{\"o}dlseder, J.,
et~al. 2005, \aap, 441, 513

\bibitem[Kobayashi et~al.(1998)]{kobayashi98} Kobayashi, C., et~al. 1998, \apjl, 503, 155L

\bibitem[Kobayashi et~al.(2007)]{kobayashi07} Kobayashi, C., Springel,
V., \& White, S.~D.~M. 2007, \mnras, 376, 1465

\bibitem[Kudritzki \& Puls(2000)]{kudritzki00} Kudritzki, R.~P., \& Puls, J. 2000, \araa, 38, 613 

\bibitem[Lee et~al.(2006)]{lee06} Lee, H., et~al. 2006, \apj, 647, 970

\bibitem[Li et~al.(2007)]{li07} Li, W., et~al. 2007, \apjl, 661, 1013L

\bibitem[Lupton(2005)]{lupton05} Lupton, R. 2005 \\
(http://www.sdss.org/dr4/algorithms/sdssUBVRITransform.htm)

\bibitem[Maeder \& Meynet(2004)]{maeder04} Maeder, A., \& Meynet, G. 2004, \aap, 422, 225

\bibitem[Mannucci et~al.(2005)]{mannucci05} Mannucci, F., et~al. 2005, \aap, 433, 807

\bibitem[MacFadyen \& Woosley(1999)]{macfadyen99} MacFadyen, A.~I., \&
Woosley, S.~E. 1999, \apj, 524, 262

\bibitem[Meynet et~al.(2006)]{meynet06} Meynet, G., Mowlavi, N., \&
Maeder, A.  2006, astro-ph/0611261

\bibitem[Mikuz(2007)]{mikuz07} Mikuz, H. 2007, CBET~933

\bibitem[Modjaz et~al.(2007)]{modjaz07} Modjaz, M., et~al. 2007, \aj,
submitted (astro-ph/0701246)

\bibitem[Neill et~al.(2006)]{neill06} Neill, J.~D., et~al. 2006, \aj,
132, 1126

\bibitem[Palacios et~al.(2005)]{palacios05} Palacios, A., et~al. 2005,
\aap, 429, 613

\bibitem[Pastorello et~al.(2007)]{pastorello07} Pastorello, A.,
et~al. 2007, Nature, 447, 829

\bibitem[Prantzos \& Diehl(1996)]{prantzos96} Prantzos, N., \& Diehl,
R. 1996, \physrep, 267, 1

\bibitem[Prantzos \& Boissier(2003)]{prantzos03} Prantzos, N., \&
Boissier, S. 2003, \aap, 406, 259

\bibitem[Prantzos(2004)]{prantzos04} Prantzos, N. 2004, \aap, 420, 1033

\bibitem[Prieto et~al.(2007)]{prieto07} Prieto, J.~L., et~al. 2007,
\aj, submitted  (arxiv:0706.4088)

\bibitem[Petrosian et~al.(2005)]{petrosian05} Petrosian, A.,
et~al. 2005, \aj, 129, 1369

\bibitem[Podsiadlowski et~al.(2006)]{podsiadlowski06} Podsialdlowski, P.,
et~al. 2006, astro-ph/0608324
 
\bibitem[Quimby et~al.(2006)]{quimby06} Quimby, R., et~al. 2006, \apj,
636, 400

\bibitem[Quimby et~al.(2007)]{quimby07} Quimby, R., et~al. 2007, CBET~927 

\bibitem[Richardson et~al.(2002)]{richardson02} Richardson, D., et~al. 2002, \apj, 123, 745

\bibitem[R{\"o}pke et~al.(2006)]{roepke06} R{\"o}pke, F.~K., et
al. \aap, 453, 203

\bibitem[Sako et~al.(2005)]{sako05} Sako, M., et~al. 2005, astro-ph/0504455

\bibitem[Scannapieco \& Bildsten(2005)]{scannapieco05} Scannapieco, E.,
\& Bildsten, L. 2005, \apjl, 629, 85L

\bibitem[Schlegel et al.(1998)]{sfd98} Schlegel, D.~J., Finkbeiner,
D.~P., \& Davis, M.  1998, \apj, 500, 525

\bibitem[Soderberg et~al.(2006)]{soderberg06} Soderberg, A.,
et~al. 2006, 638, 930

\bibitem[Stanek et~al.(2003)]{stanek03} Stanek, K.~Z., et~al. 2003,
\apjl, 591, 17L
 
\bibitem[Stanek et~al.(2006)]{stanek06} Stanek, K.~Z., et~al. 2006,
\actaa, 56, 333

\bibitem[Strolger et~al.(2002)]{strolger02} Strolger, L.-G.,
et~al. 2002, \aj, 124, 2905

\bibitem[Timmes et al.(2003)]{timmes03} Timmes, F.~X., Brown, E.~F., \&
Truran, J.~W. 2003, \apjl, 590, L83

\bibitem[Tremonti et~al.(2004)]{tremonti04} Tremonti, C.~A., et~al.
2004, \apj, 613, 898

\bibitem[Tsvetkov et~al.(2004)]{tsvetkov04} Tsvetkov, D.~Y., Pavlyuk,
N.~N., \& Bartunov, O.~S. 2004, AstL, 30, 729

\bibitem[Tully et~al.(1998)]{tully98} Tully, R.~B., et~al. 1998, \aj, 115, 2264

\bibitem[Umeda et~al.(1999)]{umeda99} Umeda, H., et~al. 1999, \apjl, 522, L43 

\bibitem[van den Bergh(1997)]{vandenbergh97} van den Bergh, S. 1997,
\aj, 113, 197

\bibitem[van der Heyden et~al.(2004)]{vanderheyden04} van der Heyden,
K.~J., Bleeker, J.~A.~M. \& Kaastra, J.~S. 2004, \aap, 421, 1031

\bibitem[Vink \& de Koter(2005)]{vink05} Vink, J.~S., \& de Koter, A. \aap,
442, 587

\bibitem[Whelan \& Iben(1973)]{whelan73} Whelan, J., \& Iben, I.~J. 1973, 186, 1007 

\bibitem[Yoon \& Langer(2005)]{yoon05} Yoon, S.~-C., \& Langer, N. 2005,
\aap, 443, 643

\bibitem[Zaritsky et~al.(1994)]{zaritsky94} Zaritsky, D., Kennicutt,
R.~C., Jr., \& Huchra, J.~P. \apj, 420, 87

\end{thebibliography}
\end{document}